\documentclass[runningheads]{llncs}

\usepackage[T1]{fontenc}
\def\doi#1{\href{https://doi.org/\detokenize{#1}}{\url{https://doi.org/\detokenize{#1}}}}
\usepackage{enumerate}
\usepackage{graphicx}
\usepackage{url}
\usepackage{amsmath}
\usepackage{subfigure}
\usepackage[most]{tcolorbox}

\usepackage{hyperref}
\usepackage{graphicx}
\usepackage{academicons}
\usepackage{setspace, xcolor}
\usepackage{amsmath, amssymb, relsize}
\usepackage{algorithm, algorithmic, listings}
\usepackage[misc,geometry]{ifsym}
\newcommand{\orcid}[1]{\href{https://orcid.org/#1}{\includegraphics[width=8pt]{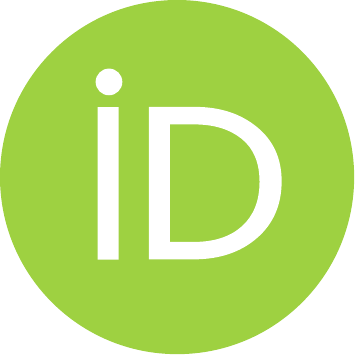}}}

\usepackage[misc]{ifsym}

\begin{document}
\title{Integrating User and Item Reviews in Deep Cooperative Neural Networks for Movie Recommendation}
\titlerunning{Movie Recommendation using Deep Cooperative Neural Networks} 
\def\bmdsc{1}

\author{%
    Aristeidis Karras\inst{\bmdsc}\orcid{0000-0002-4632-6511}
\and
    Christos Karras\inst{\bmdsc}\orcid{0000-0002-4253-7661}}

\authorrunning{A. Karras et al.} 

\institute{Computer Engineering and Informatics Department,\\ 
University of Patras, 26504 Patras, Greece\\
\email{\{akarras, c.karras\}@ceid.upatras.gr}}

\maketitle              
\begin{abstract}
User evaluations include a significant quantity of information across online platforms. This information source has been neglected by the majority of existing recommendation systems, despite its potential to ease the sparsity issue and enhance the quality of suggestions. This work presents a deep model for concurrently learning item attributes and user behaviour from review text. Deep Cooperative Neural Network (DeepCoNN) is the suggested model consisting of two parallel neural networks connected in their final layers. One of the networks focuses on learning user behaviour from reviews submitted by the user, while the other network learns item attributes from user reviews. On top, a shared layer is added to connect these two networks. Similar to factorization machine approaches, the shared layer allows latent factors acquired for people and things to interact with each other. On a number of datasets, DeepCoNN surpasses all baseline recommendation systems, according to experimental findings.

\end{abstract}

\section{Introduction}
The research on recommendation systems is extensive and, until recently, has usually centred on well-known matrix factorization methods, such as collaborative filtering ($CF$), which Netflix and Spotify have popularised. Collaborative filtering is superior than other approaches since it is computationally efficient and very simple to implement. Conversely, the disadvantages of collaborative filtering in real-world applications might provide insurmountable obstacles. To explain, CF has a "cold-start" issue since it relies on user ratings to create predictions. When consumers have little or no rating history compared to the total number of products to rate, this is known as data \textit{sparsity}. It is difficult to effectively forecast ratings/recommendations with sparse user data, resulting in poor model performance and generalisation \cite{balakrishnan2014deepplaylist}.

Recent strong empirical results in Deep Learning, particularly in natural language modelling using convolutional neural networks ($CNN$) and recurrent neural networks ($RNN$) to capture complex feature interactions in textual data, have prompted recent recommendation systems ($RecSys$) research to leverage deep learning. CNN models \cite{zheng2017joint} and RNNs \cite{catherine2017transnets}, such as \textit{Long Short-Term Memory} (LSTM)  and \textit{Gated Recurrent Units} (GRU), have shown the capacity to assist with the generalisation issue and enhance model accuracy in textual data sets.

\section{Related Work}
Our investigations concentrate on extending the $DeepCoNN$ model, thus our major reference will be the original paper \textit{Joint Deep Modeling of Users and Items Using Reviews for Recommendation} \cite{zheng2017joint}. Although the authors do not give an online repository for open-source code, we reached them via email and they gave information on model replication and suggestions for directions and improvement. In addition, we discovered a comparable code base online that leveraged data from video game reviews \cite{mcauley2015image}. Related works as $DeepFM$ \cite{guo2017deepfm}, which like our model, combines a kind of matrix factorization with a CNN architecture to capture user/item interactions; and $TransNets$ \cite{catherine2017transnets}, which extends $DeepCoNN$ to instances when the user's review is unavailable are two similar works in the field of information retrieval. We consulted \textit{Factorization Machines} from \cite{rendle2010factorization} for background information on the field.

Similar works in the field of Machine Learning as in \cite{karras2022pattern} highlight the significance of neural networks for pattern recognition while research as in \cite{karras2022dbsop} pave the wave for the integration of traditional MCMC methods with neural networks. Additionally, MCMC methods are of great importance when trying to gain useful insights from data or to approximate functions that are unable to be estimated with traditional methods \cite{karrasoverview},\cite{karrasmaximum}. Ultimately, analytics in the field of agriculture are presented in \cite{karrassaf} for knowledge extraction.

As per the initial approach for this work, it was to mimic Google's \textit{Wide and Deep} model \cite{cheng2016wide}. We spent a descent amount of time investigating this approach, but eventually concluded that we were unable to locate comparable, high-quality category data as described in the original research. Moreover, we chose to move to the $DeepConn$ model owing to its improved applicability to a given data set and the acceptance of the paper to the 2017 ACM International Conference on Web Search and Data Mining ($WSDM$). Lastly, we found the paper to be more innovative and challenging.

\section{Methodology}
This section describes the approach followed, including a summary of the problem formulation, a summary of our data collection and pre-processing stage, and model architecture information. The problem formulation is as follows. In order to make better movie recommendations how can we more accurately predict a users rating for an unseen movie based on what the user has previously seen? Furthermore, can we improve generalization of our model when information on a users past movie ratings is limited by utilizing user review data?

These two questions are the bottom line of our problem. Prior to deep learning, standard approaches for recommendation systems ($RecSys$) used collaborative filtering which relies on decomposing users, items (i.e. movies), and ratings into latent feature matrices. The interaction between users and items is captured by matrix multiplication of the weight matrices of these latent features. One common CF method includes using the cosine similarity measure between all pairs of movies that users have rated:\\

Where,
$m_{i}$ and $m_{j}$ refer to movie vectors of ratings of users who have rated both movies:

\begin{equation}
\textbf{$cos(\theta)$} = \frac{\vec{m_{i}} \times \vec{m_{j}}}{||\vec{m_{i}}||_2 \times ||\vec{m_{j}}||_2}
\end{equation}

This results in an $M \times M$ matrix of movie-to-movie similarity with ones along the diagonal. Consequently, the projected rating for movie $m_{2}$ for $user_1$ would be derived using similarity measures between ($m_2,m_1$) and ($m_2,m_3$), weighted by the ratings for $m_1$ and $m_3$.

From the above definition, it is self-evident that this strategy has a significant drawback: $sparsity$. When ratings are minimal, the movie-to-movie matrix consists mostly of zeros, reducing the capacity to anticipate. Current research initiatives, such as $DeepCoNN$, are attempting to increase the precision of sparse data by using text data that consumers provide after seeing films. This text data is utilised to train neural networks and provides extra insights compared to numerical ratings alone.

\subsection{Data Overview and Preprocessing}
The Amazon Instant Video Review 5-core data is a JSON file containing nine values per entry, of which the following four are utilised:
\begin{enumerate}[I]
\item  \textbf{ReviewerID} - \textit{user ID} 
\item  \textbf{Asin} - \textit{movie ID} 
\item  \textbf{ReviewText} - \textit{review text} 
\item \textbf{Overall} - \textit{movie rating}
\end{enumerate}
\subsection{Grouping Users and Reviews}
This paper concentrates on the DeepCoNN model formulation, which consists of two collaboratively modelled neural networks. The first network $Net_{i}$ utilises all text reviews for a particular user. The second network $Net_{p}$ utilises all text reviews for a specific item (movie). To extend the study, we create two dictionaries where each key is a unique $reviewerID$ ($Asin$) accompanied with a set of text reviews for each user and movie.

\begin{table}[htbp]
    \centering
    \caption{Amazon Instant Video Dataset Overview}
    \begin{tabular}{|l|c|c|c|c|c|c|c|}
    \hline
 & \# Reviews & \# Users & \# Movies & Training\% & Test\% \\
\hline
Data & 39.517  & 5047 & 1782 & 90\% & 10\%  \\
\hline
    \end{tabular}
    \label{tab:1}
\end{table}
\subsection{Data Preprocessing}
Using the built-in pre-processing API of Keras, each review is tokenized into distinct lowercase words without any punctuation. Then, we utilise our embedding mapper to convert the reviews into their vector representation in GloVe. Given that review durations generally vary in terms of the amount of words per review, we pad or truncate reviews such that input matrices have the same number of dimensions. As a hold-out testing set, we then exclude a subset of individuals and their reviews from the data altogether (such that their reviews do not show in the aggregate of any reviews). The dataset as well as the number of reviews, users, movies and training/testing \% split is shown in table \ref{tab:1}.

\subsubsection{Pre-trained Embeddings}

Word embeddings transform the review words of the review text into $n$-dimensional vectors. Global Vectors for Word Representation (GloVe.6B) 50-dimensional and 100-dimensional pre-trained embeddings were chosen \cite{ko2016collaborative}. GloVe is a weighted least-squares objective log-bilinear model. Intuitively, the model recognises that ratios of word-word co-occurrence probabilities are capable of storing meaning. Consequently, GloVe embeddings aid in capturing the text structure of our review data. The benefit of employing pre-trained embeddings as opposed to learning the embeddings from the data is that one may reduce model training durations while possibly improving the quality of word-vector representations in the model. The downside is that our text corpus may vary greatly from the corpus used to train GloVe.

\subsection{Model Architectures}
CNNs and RNNs are one of the most effective algorithms for a variety of NLP applications, including sentiment categorization and question answering. Given their effectiveness in text-based modelling, we focused on leveraging these models to enhance the DeepCoNN-DP model. In the subsections that follow, we describe each model type employed in our research. Beforehand, we offer a comprehensive review of the $DeepCoNN$ model described in detail in the original paper \cite{zheng2017joint}.

The architecture of $DeepCoNN$ as described in the original research article is represented here. As indicated earlier, $DeepCoNN$ employs two parallel neural networks trained concurrently. One network learns user-specific latent factor representations from reviews, while the second network learns movie-specific components.

\begin{figure}[t!]
\centering
\includegraphics[scale=0.30]{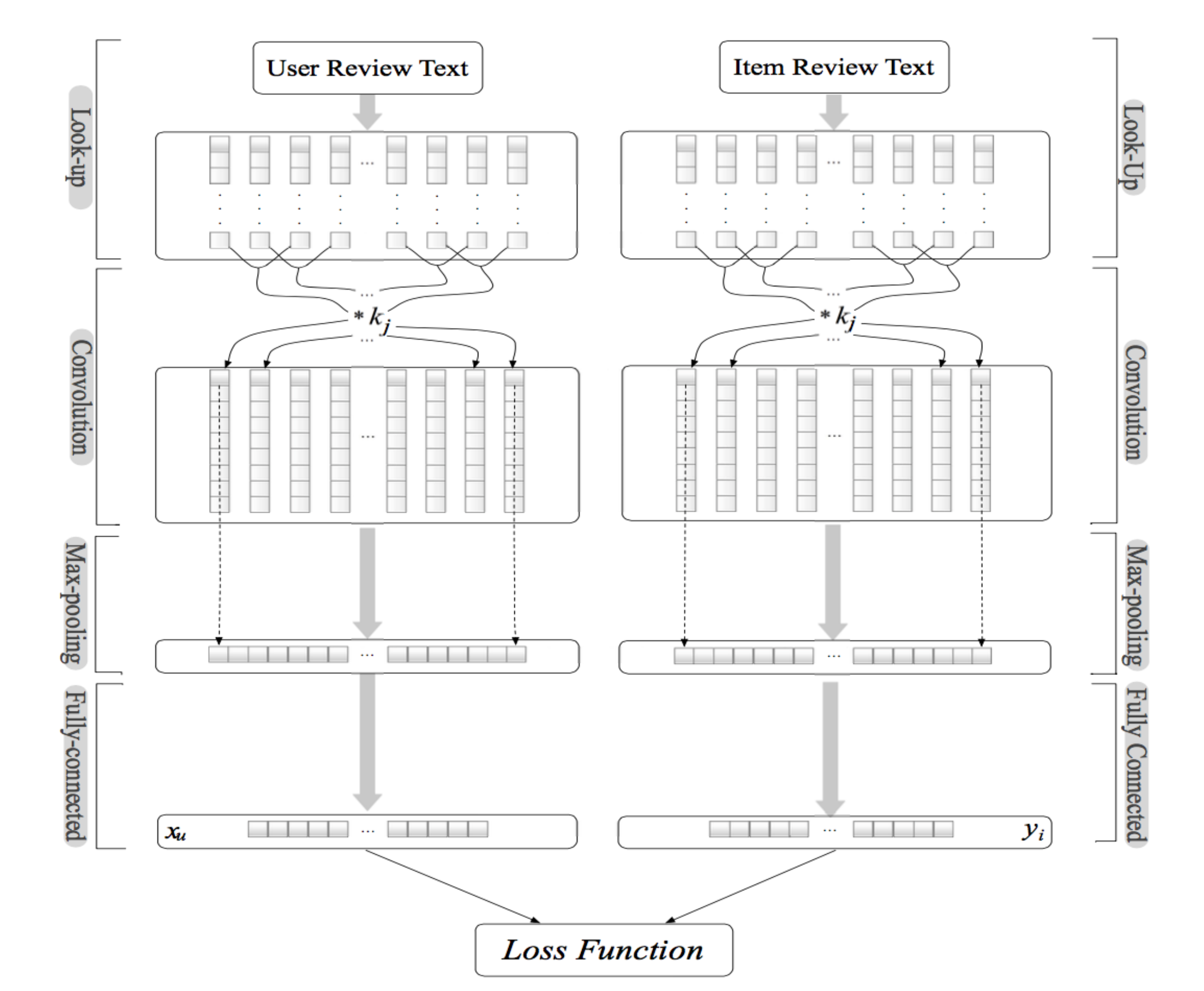}
\caption{DeepCoNN Architecture}
\label{fig:arch}
\end{figure}

Note that the first layer of the network in the original study is a "lookup" layer that translates review text into embeddings. As mentioned before, we build embedding prior to putting the data into the network, therefore we eliminate a "lookup" layer. Since the embedding layer employs pre-trained embedding, which cannot be learned, our technique is comparable. The following table, \ref{tab:arch}, provides a summary of our basic DeepCoNN-DP model.

\begin{table}[h]
\centering
\caption{DeepCoNN Baseline Architecture}
\begin{tabular}{ |p{3cm}|p{1cm}| }
 \hline
 \multicolumn{2}{|c|}{\textbf{Baseline DeepCoNN-DP Model}} \\
 \hline
 \textbf{CNN Layer} & 1  \\
 Hidden Size & 64 \\
 Filters   & 2    \\
 Kernel Size &   8\\
 Strides & 6 \\
 Activation    & ReLU \\
 \hline
 \textbf{Max Pooling} &   1  \\
   \hline
 \textbf{Flattened} & 1    \\
  \hline
 \textbf{Fully-Connected} & 1  \\
  Hidden Size & 32 \\
  \hline
\end{tabular}
 \label{tab:arch}
\end{table}

To concurrently train two networks with a single loss function, the paper concatenates the outputs of both networks. The connection between user review features and movie review features is carried out using an unspecified \textit{factorization machine} (FM). Nonetheless, the FM aims to record second-order interactions between users and films. Significantly, a factorization machine permits the capture of second order interactions with less than $N^2$ weights by expressing each interaction weight as the dot product of two lower-dimensional vectors, where each vector represents the $i^{th}$ variable. The loss function includes FM for $DeepCoNN$ and is defined as in \eqref{eq:loss}.
\begin{equation}\label{eq:loss}
DeepCoNN-FM = \hat{\beta_{0}} + \sum_{i=1}^{\hat{|z|}}\hat{w_{i}}\hat{z_{i}} + \sum_{i=1}^{\hat{|z|}}
\sum_{j=i+1}^{\hat{|z|}} \langle \hat{v_i}, \hat{v_j} \rangle \hat{z_i}\hat{z_j}
\end{equation}

Where:
\begin{itemize}
    \item $\hat{\beta_0}$ is the global bias
 \item $\hat{w_i}$ models strength of $i^{th}$ variable in $\hat{z}$
 \item $\langle\hat{v_i}, \hat{v_j} \rangle$ is the $2^{nd}$ order interaction
\end{itemize}

The authors informed us that taking the dot product of the output from the fully connected layers of $Net_{u}$ and $Net_{i}$ should roughly mirror the factorization machine technique and findings focused on first-order interactions. In the conclusions and findings section, these results are further addressed. The dot product approach loss function is defined as in \eqref{eq:dotloss}.
\begin{equation}\label{eq:dotloss}
DeepCoNN-DP = \hat{\beta_{0}} + \sum_{i=1}^{\hat{|z|}}\hat{w_{i}}\hat{z_{i}} + x^{T}_{u}x_{i}
\end{equation}

This paper optimises the original paper using $RMSprop$, which we have duplicated. However, we switched to $Adam$ in subsequent studies after determining that $Adam$ produced comparable results. As a result, we chose to employ Adam in all experimental designs to maintain model consistency. $Adam$ is an adaptive variant of gradient descent that regulates the step size in relation to the gradient's absolute value. Lastly, observe that regularisation was not mentioned in the initial paper.

\begin{table}[!ht]
\centering
\caption{DeepCoNN Architecture Comparisons}
\begin{tabular}{|l|c|c|c|c}
\hline
 {} & \textbf{CNN} & \textbf{LSTM} & \textbf{GRU}  \\
\hline
\textbf{GloVe Embedding} & 50 $|$ 100 & 50 $|$ 100 & 50 $|$ 100 \\
\hline
\textbf{Hidden Layer} & 1 & 1 & 1 \\
Units & 64 & 64 & 64 \\
Activation & ReLU & Tanh &  Tanh \\
Dropout \% & 0.10 & 0.10 & 0.10  \\
\hline
\textbf{Dense Layer} & 1 & 1 & 1 \\
Units & 64 & 64 &  64 \\
Activation & ReLU & ReLU &  ReLU \\
\hline
\textbf{Optimizer} & Adam & Adam &  Adam \\
\hline
\end{tabular}
\label{tab:comp}
\end{table}

\subsection{CNN Architecture}
CNNs are widely used in the area of image processing and its applications.
Similar to image processing, CNNs utilise temporal convolution operators known as filters for text applications. In practise, filters are employed at different resolutions and paired with non-linear activation functions and pooling strategies such as maximum pooling. Convolution is a linear mapping over $n$-gram vectors that facilitates the learning of word representations. Figure \ref{fig:arch} illustrates the architecture used in the original study.

In our first model trials, we opted to improve upon the $DeepCoNN$ model by including dropout regularisation and increasing the number of neurons in the hidden (CNN) and dense layers. In addition, we used $Adam$ as our optimizer for consistency while testing with other models. Our implementation is described in detail in the tables \ref{tab:arch} and \ref{tab:comp}.
\subsection{LSTM Architecture}
Using gated cells, Long Short-Term Memory (LSTM) cells preserve and manage information flow outside of the typical flow of a simple recurrent network (RNN). Information may be saved, written to, or retrieved from a cell symbolically, analogous to how data is processed within the memory of a computer. By opening and shutting gates, LSTM cells determine what is saved, written, or read. In reality, these operations are carried out using element-wise multiplication by sigmoids whose values range from zero to one. Table \ref{tab:arch} depicts a summary of an LSTM cell based on \cite{cho2014learning}. An illustration of our implementation is provided in \ref{fig:arch}.

Thus gates can block or pass on information based on its strength, which they control through their own sets of weights. LSTMs are well-known variants of traditional RNNs and were introduced by Hochreiter and Schmidhuber in 1997 \cite{hochreiter1997long}. Furthermore, they are popular with the natural language processing given their above abilities, in addition to, their ability to counteract the vanishing-gradient problem; and because standard stochastic gradient descent-based learning techniques can be used given their differentiability. The downside of LSTMs is the magnitude of parameterizations increases leading to longer training times, which we discuss in the results section. As discussed earlier, in all of our experiments we implemented regularization via recurrent dropout and/or regular dropout. We also expanded the number of hidden units and changed the optimizer.

Consequently, gates may block or pass information dependent on its strength, which they manage through their own sets of weights. Moreover, they are prominent in natural language processing because to the aforementioned qualities, as well as their capacity to combat the vanishing-gradient issue, and because normal stochastic gradient descent-based learning approaches may be used due to their differentiability. The size of LSTM parameterizations rises, resulting in longer training periods, as discussed in the results section. As stated before, in all of our trials, regularisation was achieved by recurrent dropout and/or regular dropout. Additionally, we increased the number of hidden units and modified the optimizer.

\subsection{GRU Architecture}
We also studied Gated Recurrent Units (GRU) which were developed in 2014 by Cho et al. \cite{goldberg2016primer}. GRUs simplify the LSTM design by combining the forget and input gates into one update gate and integrating the cell state with the hidden state. Thus GRU have fewer gates than LSTMs and remove the distinct memory component. As you might anticipate, this leads to fewer parameterizations. Below is a summary of the mathematics of GRU gates \cite{hochreiter1997long}. A overview of our implementation may be seen in table \ref{tab:comp} above.

\begin{equation}
\begin{aligned}
z = \sigma(x_t U^z + s_{t-1}W^z) \\
r = \sigma(x_t U^r + s_{t-1}W^r) \\
h = tanh(x_t U^h + (s_{t-1} * r) W^h) \\
s_t = (1-z) * h + z * s_{t-1}
\end{aligned}
\end{equation}

\begin{equation}
\begin{aligned}
\mathbf{s}_{\mathbf{j}}=R_{G R U}\left(\mathbf{s}_{\mathbf{j}-1}, \mathbf{x}_{\mathbf{j}}\right) &=(\mathbf{1}-\mathbf{z}) \odot \mathbf{s}_{\mathbf{j}-1}+\mathbf{z} \odot \mathbf{h} \\
\mathbf{z} &=\sigma\left(\mathbf{x}_{\mathbf{j}} \mathbf{W}^{\mathbf{x z}}+\mathbf{h}_{\mathbf{j}-\mathbf{1}} \mathbf{W}^{\mathbf{h z}}\right) \\
\mathbf{r} &=\sigma\left(\mathbf{x}_{\mathbf{j}} \mathbf{W}^{\mathbf{x r}}+\mathbf{h}_{\mathbf{j}-1} \mathbf{W}^{\mathbf{h r}}\right) \\
\mathbf{h} &=\tanh \left(\mathbf{x}_{\mathbf{j}} \mathbf{W}^{\mathbf{x h}}+\left(\mathbf{h}_{\mathbf{j}-1} \odot \mathbf{r}\right) \mathbf{W}^{\mathrm{hg}}\right) \\
\mathbf{y}_{\mathbf{j}}=O_{L S T M}\left(\mathbf{s}_{\mathbf{j}}\right) &=\mathbf{s}_{\mathbf{j}}
\end{aligned}
\end{equation}

\begin{equation}
\mathbf{s}_{\mathbf{j}} \in \mathbb{R}^{d_{h}}, \mathbf{x}_{\mathbf{i}} \in \mathbb{R}^{d_{z}}, \mathbf{z}, \mathbf{r}, \mathbf{h} \in \mathbb{R}^{d_{h}}, \mathbf{W}^{\mathbf{x}} \in \mathbb{R}^{d_{x} \times d_{h}}, \mathbf{W}^{\mathbf{h}} \in \mathbb{R}^{d_{h} \times d_{h}}
\end{equation}
A single gate $(r)$ controls access to the prior state $\mathbf{s}_{\mathbf{j}-1}$ and computes a suggested update $\mathbf{h}$. The updated state $\mathbf{s}_{\mathbf{j}}$ (which also acts as the output $\mathbf{y}_{\mathbf{j}}$ ) is then calculated using an interpolation of the previous state $\mathbf{s}_{\mathbf{j}-1}$ and the proposal $\mathbf{h}$, with the proportions of the interpolation controlled by the gate $\mathbf{z}$.

\section{Experimental Results}
To implement the DeepCoNN-DP model we used $Python$. Modeling construction was done via $Keras$ version 2.7 with a $TensorFlow$ version 2.6 back end. Data pre-processing was done mainly via $pandas$, $numpy$, \textit{Keras preprocessing API}, and $sklearn$. For training we used one Tesla K80 GPU on Google Cloud. Our preliminary work consisted of implementing DeepCoNN to model text review data for movie rating prediction. We then tried to innovate upon the DeepCoNN framework by experimenting with two new architectures in the $LSTM$ DeepCoNN-DP model and $GRU$ DeepCoNN-DP models as outlined above. Additionally, experimentation was done with the original CNN underlying architecture through regularization, hidden layer dimensionality, optimizers, and GloVe word embeddings. Below we present and summarize those results.

We used our recreation of DeepCoNN-DP as the baseline model when comparing experiments. Mean squared error (MSE) was used for models comparisons. Table \ref{tab:1} shows our training/testing split for the data pre-processing step.
$$
\mbox{MSE} = \frac{1}{N}\sum_{n=1}^{N}(r_{n} - \hat{r}_{n})^2
$$

In the original paper \cite{zheng2017joint} the Amazon Music Instant Review data set was used. We have included a table in \ref{tab:dataset} showing the original results of the authors from that experiment for convenience. Note that they achieved an MSE of 1.253 on the Amazon Music data set for the \textit{DeepCoNN-DP} model and a MSE of 1.233 on the \textit{DeepCoNN-FM} model.

\begin{table}[htbp]
    \centering
    \caption{Comparing model versions that have been suggested. The finest outcomes are highlighted in bold.}
    \begin{tabular}{|c|c|c|c|}
\hline \textbf{Model} & \textbf{Yelp} & \textbf{Amazon Music Instruments} & \textbf{Beer} \\
\hline DeepCoNN-User & $1.5771$ & $1.3734$ & $0.2921$ \\
\hline DeepCoNN-Item & $1.5782$ & $1.3727$ & $0.2964$ \\
\hline DeepCoNN-TFIDF & $1.7134$ & $1.4692$ & $0.5891$ \\
\hline DeepCoNN-Random & $1.7991$ & $1.5171$ & $0.6277$ \\
\hline DeepCoNN-DP & $1.4911$ & $1.2534$ & $0.2781$ \\
\hline DeepCoNN & $\mathbf{1 . 4 4 11}$ & $\mathbf{1 . 2 3 34}$ & $\mathbf{0 . 2 7 32}$ \\
\hline
\end{tabular}
    \label{tab:dataset}
\end{table}

From our experiments using the Amazon Movie Review data set we found that many experiments improved on these scores significantly. Our results from all architectures use the dot product  (DeepCoNN-DP) to capture user/movie interactions instead of the final model of the initial paper (DeepCoNN-FM). After contacting the authors of the paper they recommended to use the dot product to capture interactions since the factorization machine method omitted specifications in the paper.

\begin{table}[!h]
\centering
\caption{Results - 50d GloVe Model Comparisons}
\begin{tabular}{|l|c|c|c|c|}
\hline
 & \textbf{Embedding} & \textbf{Training Time} & \textbf{MSE Loss} \\
\hline
DC-DP & 50d & 0 hr 12 min 53 s & 1.4851 \\
\textbf{GRU}& \textbf{50d}& \textbf{1 hr 33 min 27 s} & \textbf{1.0787} \\
LSTM & 50d & 1 hr 54 min 30 s & 1.5392 \\
\hline
\end{tabular}
\label{tab:2}
\end{table}

\begin{table}[!h]
\centering
\caption{Results - 100d GloVe Model Comparisons}
\begin{tabular}{|l|c|c|c|c|}
\hline
 & \textbf{Embedding} & \textbf{Training Time} & \textbf{MSE Loss} \\
\hline
\textbf{DC-DP} & \textbf{100d} & \textbf{0 hr 17 min 45 s} & \textbf{0.8549} \\
GRU & 100d & 1 hr 49 min 52 s & 1.1249 \\
LSTM & 100d & 2 hr 3 min 46 s & 1.3328 \\
\hline
\end{tabular}
\label{tab:3}
\end{table}

\begin{table}[!h]
\centering
\caption{Results - 50d GloVe Model With Dropout Comparisons}
\begin{tabular}{|l|c|c|c|c|}
\hline
 & \textbf{Embedding} & \textbf{Training Time} & \textbf{MSE Loss} \\
\hline
DC-DP & 50d & 0 hr 12 min 1 s & 1.1379 \\
GRU & 50d & 1 hr 28 min 49 s & 1.2174 \\
\textbf{LSTM} & \textbf{50d} & \textbf{2 hr 4 min 7 s} & \textbf{1.1107} \\
\hline
\end{tabular}
\end{table}

\begin{table}[!h]
\centering
\caption{Results - 100d GloVe Model With Dropout Comparisons}
\begin{tabular}{|l|c|c|c|c|}
\hline
 & \textbf{Embedding} & \textbf{Training Time} & \textbf{MSE Loss} \\
\hline
\textbf{DC-DP} & \textbf{100d} & \textbf{0 hr 19 min 12 s} & \textbf{1.1216} \\
GRU & 100d & 1 hr 43 min 7 s & 1.8291 \\
LSTM & 100d & 2 hr 0 min 20 s & 1.4741 \\
\hline
\end{tabular}
\end{table}
\subsection{Architecture Experiments}
Testing two novel network topologies, an LSTM network and a GRU network, was our first innovation beyond the outlined DeepCoNN paradigm. There was no obvious victor amongst the three networks. In the case of low-dimensional embeddings, the GRU outperformed the other networks, but failed to improve as embedding dimensionality grew. After adding regularisation, notably a dropout layer to prevent overfitting, we discovered that the LSTM model performed the best. At a high embedding dimensionality, the DC-DP network got the lowest MSE score of $0.8549$. Then, after adding regularisation, we discovered that the DC-DP model had the highest performance, but a poorer MSE than when regularisation was not included.

The disadvantage of the RNN architectures, notably the LSTM, was that their training periods were much slower than the CNN-based DeepCoNN-DP. In table \ref{tab:2}, we can see that training the LSTM took 22\% longer than training the GRU and almost \textbf{10} times longer than training the DC-DP using one Tesla K80 GPU. CNN was consistently around one order of magnitude quicker to train than recurrent networks. Consequently, additional computational resources and effort are required to train these RNN networks compared to the CNN baseline. Hence, this may restrict its applicability in the real world for businesses with limited resources or latency issues. Lastly, we conducted naive experiments with several hyperparameters, including optimizer learning rates and dropout percent, but found that adjusting resulted to only minimal changes and that a more robust hyperparameter search was required if more time was available. 
\subsection{Embedding Experiments}
As noted before, employing GloVe embeddings offers both benefits and cons, but we judged that the positives outweighed the problems. Using pre-trained embeddings might intuitively assist with the cold start issue that happens in textual models when attempting to first learn a vocabulary from a fresh corpus. After applying our novel designs, we innovated further by comparing the performance of 50d and 100d GloVe embeddings. GloVe 100d boosts the token size from about 400.000 to approximately 1.5 million. While expanding the vocabulary may assist the model in capturing word representations for previously unknown terms, this benefit may be diminished if the movie review corpus greatly varies from the movie review text corpus. As seen in table \ref{tab:3}, the results demonstrate that increasing the token size enhanced performance. In the CNN DeepCoNN-DP, we obtained the lowest MSE score of $0.8549$.
\subsection{Regularization}
In all designs, dropout regularisation was tested. While the original research did not employ regularisation in the CNN's baseline design, LSTMs/GRUs have a propensity to quickly overfit, therefore regularisation is usually recommended. Below, we display loss curves for all of our 100d GloVe-embedded designs. For the LSTM and GRU designs, we used recurrent dropout (removal of layers) and regular dropout, which consists of concealing the activity of certain nodes. Even with regularisation, both the LSTM and GRU networks begin to overfit halfway through training, as seen by the results. Experiments revealed that a moderate amount of regularisation helped to flatten the loss curve and reduce overfitting. For the CNN architecture, we saw less over-fitting, which may explain why regularisation details were excluded from the original study. The overall results are shown in \ref{tab:vali}.

\begin{table}[htbp]
    \centering
     \caption{Validation Loss Using 100 Dimensional Embeddings and No Regularization.}
    \begin{tabular}{|l|c|c|c|}
    \hline
    \textbf{Method} & \textbf{Loss} & \textbf{Validation Loss} & \textbf{MSE}\\
    \hline
    CNN 100 & 0.8732 & 1.5669 & 0.8549\\
    CNN Dropout 100 & 0.8342 & 1.4244 & 1.1216\\
    CNN Dropout & 0.7736 & 1.4842 & 1.1379\\
    CNN & 0.7163 & 1.6568 & 1.4851\\
    GRU 100 & 0.5565 & 1.7683 & 1.1246\\
    GRU Dropout 100 & 0.6590 & 1.5742 & 1.8291\\
    GRU Dropout & 0.6683 & 1.5836 & 1.2174\\
    GRU & 1.2098 & 1.7747 & 1.0787\\
    LSTM 100 & 0.5141 & 1.6888 & 1.3328\\
    LSTM Dropout 100 & 1.2039 & 1.7438 & 1.4741\\
    LSTM Dropout & 1.2610 & 1.6724 & 1.1107\\
    LSTM & 0.6789 & 1.5421 & 1.5392\\
    \hline
    \end{tabular}
    \label{tab:vali}
\end{table}

\newpage
\section{Discussion and Conclusions}

Several of our experimental outcomes were unexpected and need additional investigation. Our results for the convolutional model were initially very promising, so why, at a high dimensionality did our attempt to regularize the network fail, when it succeeded at a low dimensionality? While one might assume that dropout would be ineffective at improving convolutional networks because of how few parameters there are, Srivastava and Hinton have found this to be false \cite{srivastava2014dropout}. 

Similarly, in a 50 dimensional embedding we were able to improve our convolutional model accuracy through regularization. We believe that by trying to maintain comparability across our models our dropout hyperparameter was not optimized for all cases and caused a decrease in generalization. We attribute the poor performance of our 100 dimensional regularization in general to the same problem. It would be interesting to attempt to optimize the dropout hyperparameter across all three models independently, to find the best performer. Our goal, however, was to do a direct comparison across CNN, GRU, and LSTM models.

Equally unexpected was the observation that increasing the dimension of our embedding affected the accuracy of the GRU model. Overfitting in the embeddings is a potential reason for the GRU's diminished performance. Empirical evidence has indicated that low-dimensional word embeddings are sometimes, but rarely, superior to their high-dimensional counterparts \cite{arora2015latent}. Due to the difference in training corpora, we anticipate that the higher embedding dimensionality did not enhance the accuracy of our model. Rather of starting with a pre-trained embedding, it is likely that future research may train a complete embedding for the language.

In conclusion, the DeepCoNN model represents a step forward in recommendation systems and the typical collaborative and content-based filtering systems that are traditionally used in recommendation systems. This was confirmed by the results contained in this research on predicting movie rankings. Our research indicates that the DeepCoNN model can be significantly improved by utilizing different word embedding structures and regularization via dropout. 

Moreover, our research shows that incremental improvement to prediction accuracy is available by implementing the DeepCoNN model with different architectures, namely the LSTM and GRU architectures described above. The tradeoff with LSTM and GRU architectures is that they are significantly more complex and also more time consuming to train. Practitioners may well continue to focus on collaborative filtering and the standard DeepCoNN model for these reasons, however with improved computing power the LSTM and GRU architectures would certainly overcome the training time disadvantage.

We conclude that the DeepCoNN model outperforms traditional collaborative filtering strategies. Our investigation into different architectures and regularisation techniques confirms that more improvements to DeepCoNN are possible and that further study of this model is necessary.

\bibliographystyle{splncs04.bst}
\bibliography{bib.bib}
\end{document}